\documentclass[journal]{IEEEtran}

\usepackage[pdftex]{graphicx}
\usepackage{url}
\usepackage{subfigure}
\usepackage[pdftex]{graphicx}
\DeclareGraphicsExtensions{.pdf,.jpeg,.png}

\begin{document}
\title{Serious Games for Wrist Rehabilitation\\in Juvenile Idiopathic Arthritis}
\author{Fabrizia Corona$^1$, Rocco M. Chiuri$^2$, Giovanni Filocamo$^1$, Michaela Fo\`a$^1$,\\Pier Luca Lanzi$^2$\thanks{Contact autor: pierluca.lanzi@polimi.it}, Amalia Lopopolo$^1$, and Antonella Petaccia$^1$\\
$^1$Dipartimento della Donna, del Bambino e del Neonato\\
Fondazione IRCCS Cà Granda Ospedale Maggiore Policlinico, Milano, Italy.\\
$^2$Dipartimento di Elettronica, Informazione e Bioingegneria, Politecnico di Milano, Milano, Italy.
}

\maketitle

\begin{abstract}
Rehabilitation  is a painful and tiring process 
	involving series of exercises that patients must repeat over a long period. 
Unfortunately, patients often grow bored, frustrated, and lose motivation 
	making rehabilitation less effective.
In the recent years video games have been widely used to implement rehabilitation protocols
	so as to make the process more entertaining, engaging and to keep patients motivated.
In this paper, we present an integrated framework we developed for the wrist rehabilitation
	of patients affected by Juvenile Idiopathic Arthritis (JIA)
	following a therapeutic protocol at the Clinica Pediatrica G. e D. De Marchi.
The framework comprises four video games and a set modules that let the therapists tune and control 
	the exercises the games implemented, record all the patients actions, replay and analyze
	the sessions. 
We present the result of a preliminary validation we performed with four poliarticular JIA patients at the clinic 
	under the supervision of the therapists. 
Overall, we received good feedback both from the young patients,
	who enjoyed performing known rehabilitation exercises using video games,
	and therapists who were satisfied with the framework and its potentials for engaging and motivating 
	the patients.
\end{abstract}

\section{Introduction}
Physical therapy is essential for the treatment of disabling chronic or acute pathologies.
It usually involves a series of exercises that patients must repeat over 
	a long period to improve their condition, or at least not to worsen it. 
Unfortunately, patients tend to become bored, frustrated, and to lose motivation \cite{c5} 
	and this often reduces the effectiveness of rehabilitation. 
In the recent years, video games have increased in popularity also thanks to the new input devices 
	(e.g., the Wii Remote,\footnote{\url{https://en.wikipedia.org/wiki/Wii_Remote}} the Balance Board,\footnote{\url{https://en.wikipedia.org/wiki/Wii_Balance_Board}} the Leap Motion,\footnote{\url{https://en.wikipedia.org/wiki/Leap_Motion}} and the Microsoft Kinect\footnote{\url{https://en.wikipedia.org/wiki/Kinect}}) that made interacting with video games more intuitive \cite{c5}. 
This has provided therapists and researchers with new means
	to deploy rehabilitation protocols that could entertain and engage patients, 
	while keeping them motivated longer \cite{c8,c9}.
Off-the-shelf solutions have been tested in rehabilitation scenarios \cite{c10,c11,c12} 
	but they are generally infeasible in practice \cite{c13}.
Accordingly, 
	several researchers and therapists have worked on the development of 
	serious games targeting specific diseases and rehabilitation protocols.

In this work, we focused on the wrist rehabilitation 
	for in children affected by Juvenile Idiopathic Arthritis (JIA) that are patients of the
	Clinica Pediatrica G. e D. De Marchi.\footnote{\url{http://www.fondazionedemarchi.it}}
JIA comprises a set of rheumatoid diseases for which has not been defined a cause yet. 
The main common symptom is a chronic joint inflammation that
	can start before patients reach the age of sixteen and if the symptoms
	last from six weeks to three months, the disease is called chronic. 
We developed four video games, using the Leap Motion Controller, 
	implementing a set of rehabilitation exercises
	that the young patients at the Clinica Pediatrica G. e D. De Marchi routinely have to perform
	as part of their protocol.
The video games are part of an integrated framework that includes, 
	a tuning module that let the therapists specify the parameter of the exercises;
	a tracking module that records all the actions performed by the patients within the game and 
		provides raw data for analysis or replaying;
	a replay module that let the therapists review what a patient has done during the exercises;
	a module to let therapists create their own game levels (i.e., exercises);
	a module to generate random game levels within the constraints specified by the therapists.
All the video games provide positive feedback to the patients and implement a simple adaptive mechanism
	to decrease the game difficulty (or stop the game) when the exercises are not performed within
	the constraints specified by the therapist.
We present a preliminary validation of the framework
	we performed at the Clinica Pediatrica G. e D. De Marchi under the supervision
	of the therapists.

\section{Related Work}
\label{sec:related}
We present a brief overview of the research on games for the rehabilitation of the upper limbs. We refer the interested reader to 
	one of the available surveys for a broader view of the area of video games for rehabilitation (e.g., \cite{lohse,ONeil2014,Bonnechere2018,DBLP:journals/tciaig/PirovanoMBLB16}).

\subsection{Non-hands-free games}
These games require the patient to handle a physical device in order to give the input to the game. 
Cifuentes-Zapien et al. \cite{c15} developed a racing game intended for the rehabilitation of children with cerebral palsy, using a robot as input device. The player can control the car’s horizontal position by the pronation and supination motion. The goal of the game is to keep the car inside the track defined by the therapist. The player’s trajectory is recorded and analyzed by a software. The game records the player exercises and gives to the therapist a feedback about the level of the exercises (since the therapist himself defines the track). 
Godfrey \cite{c33} combined a robot designed to aid finger and thumb extension and flexion with two interactive virtual reality games to enhance user motivation in performing post-stroke rehabilitation exercises. The first game is a gate game in which subjects start in a flexed position and control two circles on the screen with finger and thumb movement. A moving wall with two open gates sweeps across the screen and the subject opens the fingers and thumb to pass each circle through its respective gate. The second game is an isometric squeeze and release exercise. The subjects’ fingers and thumb are held open at half their range of motion and two circles are displayed on the screen, representing finger and thumb force production. The goal is to bring the circles into a central channel by flexing. Successful flexion then activates a wall that sweeps across the screen. Patients must then relax their flexors to avoid hitting the wall. Both games have some defined parameters that allow to statically adapt the difficulty of the exercise to the player’s performance.
Dunne et al. \cite{c34} designed three  games to be played with a multi-touch display by children affected by cerebral palsy. The first is a simple game in which the players need to maneuver a bone to a dog using their finger, avoiding the other characters and the obstacles in the game landscape. To increase motivation, the game rewards the player with extra points both for performing specific actions in the game environment, such as passing with the finger upon a star, and for keeping an upright position. The therapist can customize the levels in order to tailor the game to each patient's motor ability. In the second game, the patient must spell the animal shown in a bubble onscreen using the letter tiles scattered on the playing surface. Points are used as reward and feedback of the performance. There is also a negative feedback in the form of a penalty if the patient’s compensatory movement exceeds the limits. Finally, the third game consists in catching butterflies with a jar. Also in this game a negative feedback is given to the player when he/she does something wrong. Furthermore the therapists can modify some parameters in order to customize the levels.
Karime et al. \cite{c16} designed a racing game for the rehabilitation of people with injured wrists,  in which the player has to challenge other cars. It uses a stress ball integrated with sensors and actuators as input device. The aim of this game is to help the patients performing wrist rehabilitation at home, given the portability of the stress ball. The patient controls the car by grasping and rotating the ball which is integrated with a pressure sensor, an accelerometer, and two vibrator-motors that give a haptic feedback to the player. The system can be configured according to the level of the player and stores the sensory data in a database that could be used later on by the therapist to track the patient’s progress. 
Burke et al. \cite{c36} used a webcam and Augmented Reality (AR) to create a version of Atari’s Breakout\footnote{\url{https://en.wikipedia.org/wiki/Breakout_(video_game)}} for post-stroke rehabilitation. There is a row of bricks at the top of the playfield and the player has to clear them by rebounding a ball with a paddle, which they control by moving a real-world physical object with an AR marker attached. They also prototyped another game where the player has to put real objects on virtual shelves. Mukai et al. \cite{DBLP:conf/robio/MukaiSAK17} developed a rehabilitation robot for hand. Cifuentes-Zapien et al. \cite{5871877} developed a video game in LabVIEW and MATLAB for the rehabilitation of the pronation and supination movements of children with cerebral palsy. 

\subsection{Hands-free games}
Ustinova et al. \cite{c17} developed a game called Octopus which aims to improve arm-postural coordination in patients with traumatic brain injury. The goal of the game is to pop the bubbles blown by an octopus either with the left or the right hand. Patient-computer interaction is obtained using a 6-camera system for motion capture and hand avatars implemented with three reflective markers attached to each hand. The game gives a reward in the form of either a score or new characters when the bubbles are intercepted, it also helps the patient having a feedback of his/her performance; the game also keeps track of the patient’s movements and analyzes his/her coordination.

Burke et al. \cite{c13} present a series of games which use different inputs. They made a game that uses magnetic sensor-based virtual reality equipment to track upper limb movements. It is a “whack a mole”-like game in which the player has to use his/her hand as a hammer and hit a mouse moving on the screen. They also developed a couple of games that use a webcam as the input and a marker (e.g. a glove) to keep track of the patient’s hands movements. Both games require the patient to intercept an object on screen with a certain timing.

Friedman et al. \cite{c35} used a customized version of Frets on Fire, an open-source music game inspired by Guitar Hero, combined with an instrumented glove and tested the effectiveness of this combination for post-stroke hand rehabilitation. 
Zhang et al. \cite{c37} created a system for post-stroke hand rehabilitation that integrates videogames, AR and an instrumented glove. The game involves a virtual piano that the patient can play with his fingers. They designed different levels of difficulty both to challenge the single patient and to take into account the different physical conditions of the patients. They also implemented a scoring module, both visual and audio feedbacks as performance indicators for the patient, and quantitative feedbacks for the therapist to analyze. 

\section{Juvenile Idiopathic Arthritis and Polyarthritis}
\label{sec:juvenile}
Juvenile Idiopathic Arthritis (JIA) identifies a set of autoimmune and inflammatory conditions that can develop in children ages sixteen and younger; if these conditions last for at least six weeks, the arthritis is considered chronic. About one child in every one thousand develops some type of juvenile arthritis \cite{c18,Alex4}. The term JIA was introduced in 1997 \cite{c19} and has largely supplanted the terms Juvenile Chronic Arthritis (JCA) and Juvenile Rheumatoid Arthritis (JRA) in referring to childhood chronic arthritis. The causes of the disease are still poorly understood but it seems to be related to both genetic and environmental factors which result in the heterogeneity of the illness \cite{Ravelli2007767}.
JIA is not considered hereditary and rarely involves more than one family member. Some individuals may have a genetic tendency to develop JIA, but the disease appears only after exposure to an infection, physical trauma or other unknown trigger. Arthritis Research UK \cite{Alex26} identifies five compact onset types for JIA (i) Systemic arthritis; (ii) Oligoarthritis; (iii) Polyarticular arthritis; (iv) Psoriatic arthritis; and (v) Enthesitis-related arthritis.

\subsection{Treatments}
A cure for chronic arthritis is yet to be found, but luckily there are many cases of spontaneous remission. Therapy indeed aims at inducing such remission 
while controlling pain and preserving range of motion, muscle strength, physical and psychological development \cite{Alex27}. 
Most children with chronic arthritis are treated using a combination of four main factors: the pharmacological management of the disease, physical therapy, nutrition, and orthopaedic surgery. Pharmacological treatment typically begins as soon as the disease is discovered since the sooner it starts, the less probable it is that there will be permanent sequelae. Nonsteroidal anti-inflammatory drugs are likely to be the first medicine used to reduce inflammation and pain \cite{Alex28}. Physical therapy is used to minimize pain, maintain and restore functionality, prevent deformity and disability, correct wrong compensatory behavior. Nutrition is an important aspect of long-term management and nutritional and vitamin supplementation are often used. Orthopaedic surgery has a limited role in management of chronic arthritis in young children. In the older children might be used for joint contractures, dislocations, or joint replacement. 

\subsection{Physical Therapy}
One of the main symptoms of the juvenile idiopathic arthritis (JIA) is joint inflammation that, if left untreated treated, can result in the loss of articular functionality and in the consequential worsening of the patient’s quality of life. Moreover, the child interiorizes incorrect compensatory postures or movements that persist even after a full recovery. These compensations burden on muscles and other joints, leading to new possible physical problems. Accordingly, the main goal of physical therapy is not to heal the inflammation, but rather to help the patients managing their symptoms and improving their self-sufficiency. The therapist guides the children in the process of understanding their moving capabilities, both in everyday and in sport activities. The therapist also aims at reducing the patients' fear of pain and the families propensity to overprotect. Physical therapy has to be customized for each patient and should take place both at the hospital (or another designated structure) and at home. 
%
%
There are exercises for large joints, such as the knees or the shoulders, and small joints, such as those of the hand. In this work, we focused on the exercises for the rehabilitation of the hand (with a specific focus on the wrist).

\subsection{Hand Rehabilitation} 
\begin{figure}
\centering
\begin{tabular}{cc}
\parbox{.5\columnwidth}{
	\includegraphics[width=.25\textwidth]{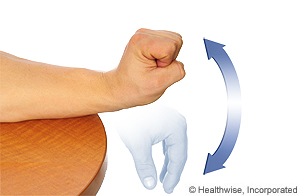}
} & 
\parbox{.5\columnwidth}{
	\includegraphics[width=.25\textwidth]{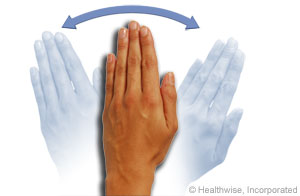}
} \\
(a) & (b) \\
\end{tabular}

\caption{Wrist exercises: (a) extension and flexion;  (b) radial and ulnar deviation.}
\label{fig:rocco3132}
\end{figure}

The wrist extension/flexion exercise is very important for JIA patients. It involves one of the two main wrist movements (Figure~\ref{fig:rocco3132}a) and it is performed either with the hand opened or closed, palm down. The forearm can be laid on a pillow or on a plane surface and it must not move during the exercise. The patient has to bend the wrist to move his/her hand upward, then lower his/her hand. Wrist deviation (Figure~\ref{fig:rocco3132}b) is the other main wrist movement and it can be ulnar o radial depending on which side the hand moves. The exercise is carried out on a perpendicular axis with respect to the flexion/extension exercise. Again, hand can be either opened or closed, palm down. The patient has to slowly bend the wrist as far as he/her can from side to side.

\section{Designing Games for Rehabilitation}
\label{sec:requirements}
Burke et al. \cite{c13} identified a set of features that a rehabilitation game should have: 
(i) it should provide precise data recording, in order for the therapist to clearly evaluate the exercise execution;
(ii) it should provide feedback to the therapists (about the current state and progress of the therapy) and to the patients
	(about their performance and progress);
(iii) it should deal with positive and negative feedback effectively in a way that keeps the patient engaged and motivated;
(iv) it should be challenging (but not frustrating) for the patient and this may be achieved statically (e.g. with an adequate level structure) or dynamically 
	(e.g. while playing) adapting the game difficulty according to the patients performance and ability.

Later, Borghese et al. \cite{c14} identified other three key features for designing rehabilitation games: adaptation, monitoring, and real-time evaluation. 
The game should adapt its difficulty level to the abilities of the player to avoid frustration (if too difficult) and boredom (if too easy to play). 
The player performance should be monitored and the game should enforce a correct execution of the rehabilitation movements. The player should receive feedback about her performance while playing to understand when she is doing something wrong or right. 
Borghese et al. \cite{c14} also remarked the importance of designing rehabilitation games according to good game design principles in order to keep the player engaged. They stated that the patient should feel like a player, focused on having fun while playing the game, rather than a patient exercising.

Nixon and Howard \cite{c31} defined a set of game design principles useful to create engaging rehabilitation games. 
Firstly,  they argued that an engaging story or context is crucial when trying to draw players into a play scenario. 
Secondly, the user interface should be intuitive and easy to understand since players should focus on how to beat the game, not how to learn its interface. 
Similarly to Burke \cite{c13} and Borghese \cite{c14}, also Nixon and Howard \cite{c31} remark the importance of providing immediate feedback 
	to the player about what she is doing right or wrong. 
Finally, players should be rewarded to keep them engaged in the game and to encourage them to continue to play (that is, to continue the therapy) 
	so as to improve their skills and rewards.

Mader et al. \cite{c32} noted that challenge and variability are two fundamental features tightly related to the design of therapeutic games.
To keep patients motivated, games should keep them updated about their progression towards their therapeutic goals. 
Games should also adapt the challenge level to prevent anxiety or boredom. 
Variability accounts for the motivation in the long run and patients should be faced with new patterns to learn, new information, 
	and new strategies to try.

\section{Video Games for Wrist Rehabilitation in JIA}
\label{sec:design_wrist}
We developed an integrated framework for wrist rehabilitation of children affected by JIA 
	that are following a therapeutic protocol at Clinica Pediatrica G. e D. De Marchi.
Our framework takes into account the requirements and desiderata we received from the therapists as well as the  
	features suggested in the literature \cite{c13,c14,c31,c32}. 

\subsection{Preliminary Meeting and Requirements}
\begin{figure}
	\centering
	\begin{tabular}{cc}
	\parbox{.45\columnwidth}{
	\includegraphics[width=.45\columnwidth]{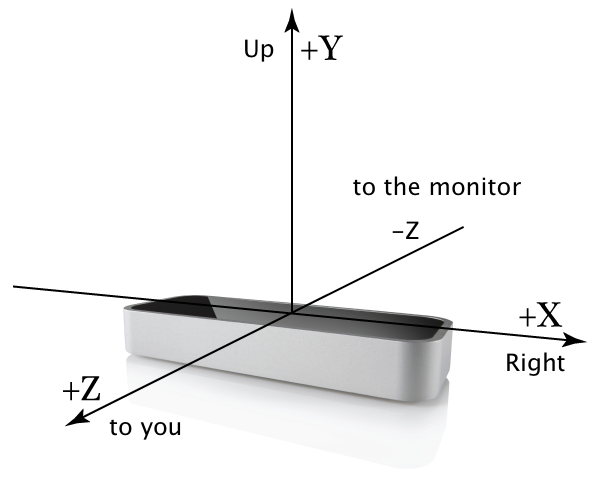}
	} & 
	\parbox{.45\columnwidth}{
	\includegraphics[width=.45\columnwidth]{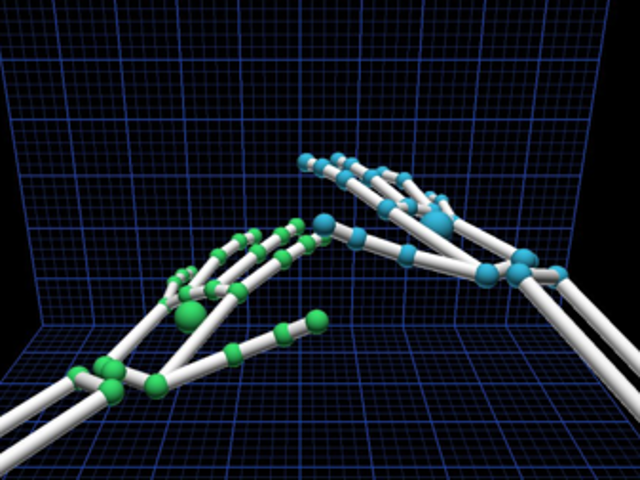}
	} 
	\\
		(a) & (b) \\
\end{tabular}
		
\caption{Leap Motion Controller: (a) the device; (b) the perceived input.}
\label{fig:leap}
\end{figure}

At first, we met with the therapists who briefly introduced JIA and demonstrated a set of rehabilitation exercises  with patients
	focused on wrist flexion, wrist extension, and radial/ulnar deviation.
We suggested the therapists the use  of Leap Motion and four game mechanics that would implement the proposed rehabilitation exercises.
Leap motion uses optical sensors and infrared light. 
The sensors are directed along the y-axis – upward when the controller is in its standard operating position (facing upward) and have a field of view of about 150 degrees (Figure \ref{fig:leap}a). 
The Leap Motion software combines its sensor data with an internal model of the human hand to help cope with challenging tracking conditions
	(Figure \ref{fig:leap}b).
	
Therapists asked to record the data about the patients hand movements and to have a quantitative feedback from the games
	(e.g., the wrist’s extension, flexion and deviation degrees throughout the game). 
Therapists also asked to record more general statistics (e.g., how long and how frequently patients played the games). 
Finally, they asked to be able to tune the games for specific patients both in terms of range of motions allowed and in length and difficulty of a game.

\begin{figure}
	\includegraphics[width=\columnwidth]{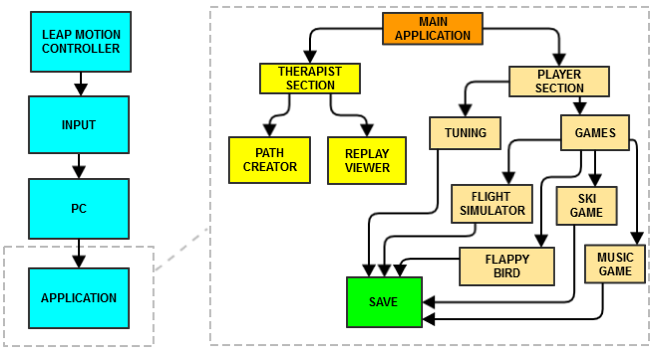}
\caption{Architecture of the integrated system.}
\label{fig:architecture}
\end{figure}


\begin{figure}
\begin{tabular}{cc}
\parbox{.48\columnwidth}{
	\includegraphics[width=.48\columnwidth]{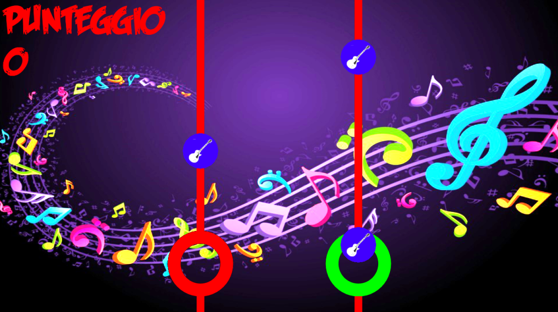}
} & 
\parbox{.48\columnwidth}{
	\includegraphics[width=.48\columnwidth]{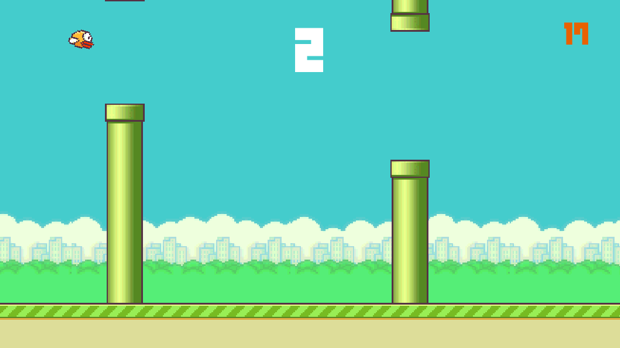}
} \\
(a) & (b) \\
\parbox{.48\columnwidth}{
	\includegraphics[width=.48\columnwidth]{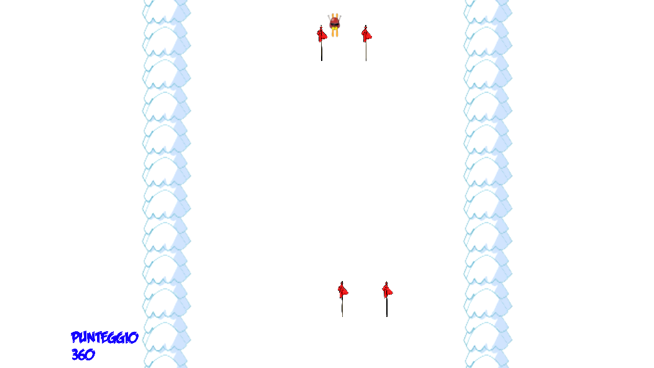}
} & 
\parbox{.48\columnwidth}{
	\includegraphics[width=.48\columnwidth]{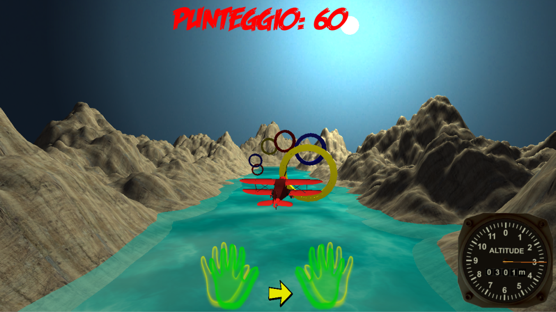}
} \\
(c) & (d) \\
\end{tabular}
\caption{The four rehabilitation games: (a) rhythm game; (b) flappy bird clone; (c) skiing; (d) plane simulator.}
\label{fig:games}
\end{figure}

\subsection{System Architecture}
Figure ~\ref{fig:architecture} shows the architecture of our framework. 
There are two main modules one of the therapists and one for the players (i.e., the patients). 
The former let the therapists create content by editing new levels for one of the available games (using the leap motion)
	and replay saved sessions.
The patient module let the therapist tune the protocol for a specific subject and give access to the rehabilitation games to the patients. 

\subsection{Rehabilitation Games}
Children affected by JIA are boys and girls of substantial age difference.
Accordingly, we developed games that could appeal such a wide range of patients and focused 
	the design on four casual games of different genres: a musical game (with a mechanic similar to Guitar Hero\footnote{\url{https://en.wikipedia.org/wiki/Guitar_Hero}}), a casual game, (a flappy-bird clone\footnote{\url{https://en.wikipedia.org/wiki/Flappy_Bird}}), an arcade game, and a 3D game (a simplified flight simulator). These games have an  intuitive gameplay so that the patient can quickly play the game without long training sessions 
	(as suggested in \cite{c31}). We also designed a balanced reward system that rewards a correct performance but, at the same time, does not penalize too much the possible errors. 

\medskip\noindent\textbf{Rhythm game.}
The aim of this game is to help patients doing wrist flexion exercises. While music plays in the background, buttons fall from above and the player has to push them at the right time. Figure \ref{fig:games}a shows a screenshot of the game: there are two lines, one for each hand; the buttons come down on each line and the player has to press them with the correct hand while they are in the corresponding circle. The pressing is achieved by the flexion motion, keeping the forearm still, horizontal, and quickly bending down the wrist, as if the player is trying to push a physical button like the ones in quiz shows.

\medskip\noindent\textbf{Flappy Bird Clone.}
This game targets both wrist flexion and extension. 
The player controls a bird that must pass through a series of tubes (Figure \ref
	{fig:games}b) until the end of the track. The game is played with one hand at a time. 
We implemented two ways of moving the character, one involving only the extension movement, the other involving both extension and flexion movements. The first game mode is similar to the original Flappy Bird game. In the original game, the player has to tap on the screen in order to make the character jump at the correct height to pass through the pipes. 
This action is recreated in our game using a movement similar to the one used in the rhythm game to press the buttons, but in the opposite direction. The player has to quickly bend up the wrist to make the character jump, as if she was trying to  create a gust of wind beneath the little bird.
%
In the second game mode the player can directly control the character’s height using his wrist motion. Keeping the hand parallel to the ground will make the character stand in the center of the screen. Bending up the wrist will make the character go up, while bending the wrist down will make it go down. The height reached by the character depends on the motion degree of the wrist. 

\medskip\noindent\textbf{Skiing Game.}
This game involves both wrist extension/flexion and radial/ulnar deviation, although not in the same exercise. The player controls a skier descending a slalom track and has to pass through the gates 
	(Figure \ref{fig:games}c).
The player can move the character right or left using one hand either 
	(i)  by radial and ulnar deviation, with the palm facing down, or 
	(ii) by wrist extension and flexion, rotating the hand by 90 degrees and making a movement like a slap. 	
As for the previous game, the wider the movement, the further the characters moves on the screen. 
Tracks can be either randomly generated based on parameters set by the therapist (e.g. duration, number of flags, range of motion) or created by the therapist. 

\medskip\noindent\textbf{Plane Simulator.} This game targets both wrist
	flexion/extension and radial/ulnar deviation in the same exercise. 
	It can be played either with one or two hands that can be opened or closed to a fist.
The player pilots a plane through a series of rings with the hands mimicking the movements of the plane itself (Figure \ref{fig:games}d). 
Bending up the wrist will tilt the plane upwards while bending the wrist down will result in the plane tilting downwards. Same goes for bending the wrist to the left and to the right.
During the testing sessions we observed that children tend to immerse in the game, forgetting to control their hands positions, so we added some visual help for both the two hand mode and the single hand mode.
In the two hands mode the main problem was that the children were often overlapping their hand, resulting in bad tracking). We added  two placeholders (the yellow hand shapes in Figure Figure \ref{fig:games}d) for the hands that indicate the optimal distance between the hands. Two hands overlays (the green hand shapes in Figure Figure \ref{fig:games}d) move accordingly to the player’s hands and show the real distance between them. If the hands are not too close or too far apart, the overlays are colored green; otherwise they become red and a warning message appears on the screen.

\section{Experiments}
\label{sec:experiments}
We performed a set of experiments with patients 
	to tune the games and to get a preliminary evaluation of our framework.
Four poliarticular JIA patients took part to the experiments which were structured over four sessions.
Subject S1 was ten years old, S2 was fifteen years old, S3 was twenty one years old, 
	S4 was twenty years old.
S1 had limitations to the ankle movements and was introduced into the experiment
	to test whether the Leap Motion could be used also to track feet movements;
S2 and S3 had articular issues affecting their hands and wrists.	
S4 had movement limitations affecting the hand and wrist joints.


\begin{figure}
\begin{tabular}{cc}
\parbox{.48\columnwidth}{
	\includegraphics[width=.48\columnwidth]{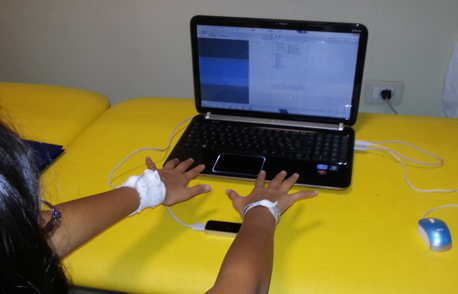}
} & 
\parbox{.48\columnwidth}{
	\includegraphics[width=.48\columnwidth]{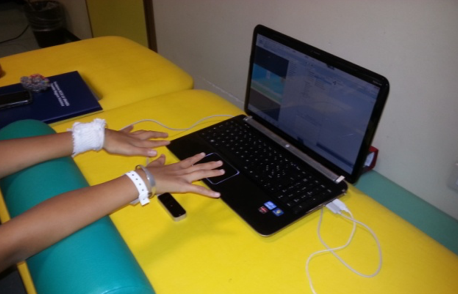}
} \\
(a) & (b) \\
\parbox{.48\columnwidth}{
	\includegraphics[width=.48\columnwidth]{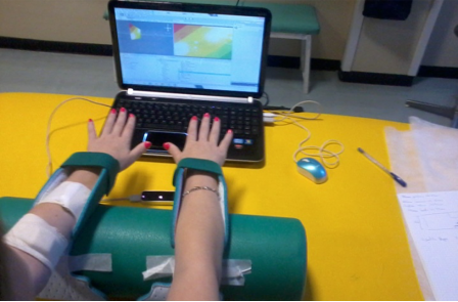}
} & 
\parbox{.48\columnwidth}{
	\includegraphics[width=.48\columnwidth]{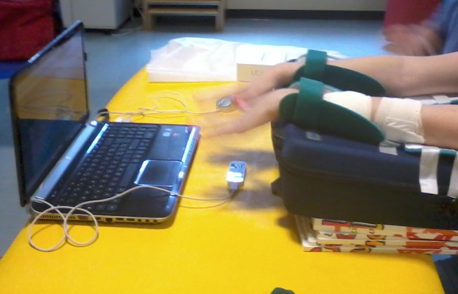}
} \\
(c) & (d) \\
\end{tabular}
\caption{The four experimental setups tested: 
	(a) without a bolster; 
	(b) with a bolster; (c) with the bolster and the orthoses; (d) with the wedge and the orthoses.}
\label{fig:configuration}
\end{figure}

\subsection{Setup}	
Each session took place at the Clinica Pediatrica G. e D. De Marchi and 
	was run on a notebook HP with a 15.6" screen. 
The notebook was placed either on a desk or on a medical bed, based on the therapists choice. 
The Leap Motion controller was placed between the notebook and the subject, 
	on the same surface as the notebook, facing upward, about 12 cm away from the notebook.
In front of the notebook was placed a chair with a seat that would let 
	the subjects extend their arms with the hands above the Leap Motion, 
	keeping a distance of at least 10 cm between the hands and the controller.

We tested four different configurations for the positioning of the arms. 
One had the subjects keeping their arms above the leap motion without any support (Figure \ref{fig:configuration}a). 
The second one added a bolster under the forearm, as a support (Figure \ref{fig:configuration}b). 
These two settings had the subjects keep their forearms as straight as possible, moving only the wrists in order to play. The second configuration helped reducing the effort of the subjects shoulders. 
We opted for this setup because the bolster is frequently used in physical therapy, 
	hence it is often available in a dedicated ward.
The third configuration (Figure \ref{fig:configuration}c) 
	added a couple resting orthoses on the bolster 
	to avoid the compensatory movement of the forearms.
The fourth configuration (Figure \ref{fig:configuration}d) 
	replaced the bolster with a higher wedge since 
	we noticed that subjects hands went too close to the controller when performing a flexion movement.
The support was also changed because the therapists noticed the subjects tendency to tilt her forearms back and forth, hence the need of a wider surface to lay the arms.

\subsection{Experimental Sessions}
\begin{figure}
	\includegraphics[width=\columnwidth]{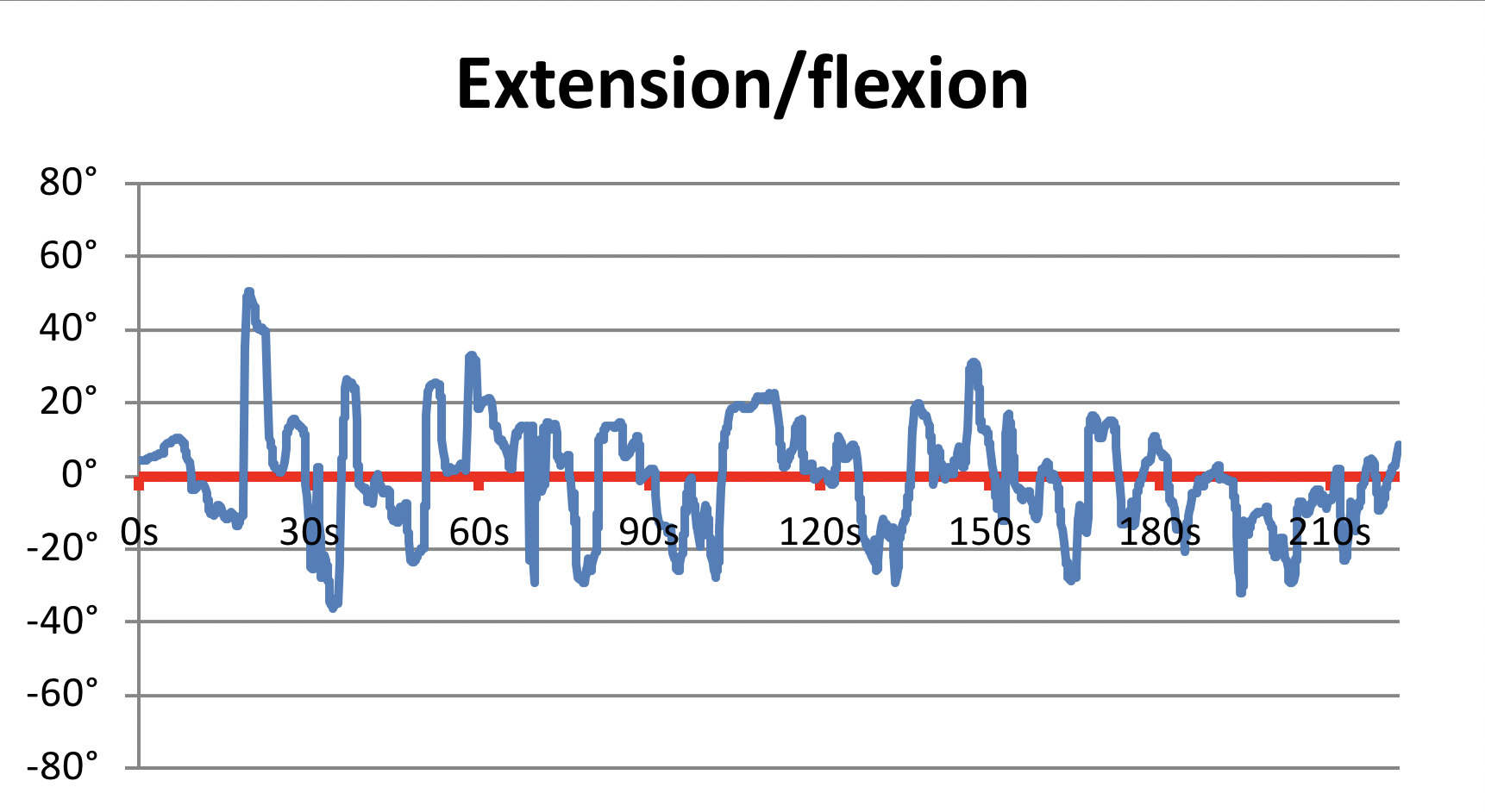}

	\centerline{(a)}
	
	\includegraphics[width=\columnwidth]{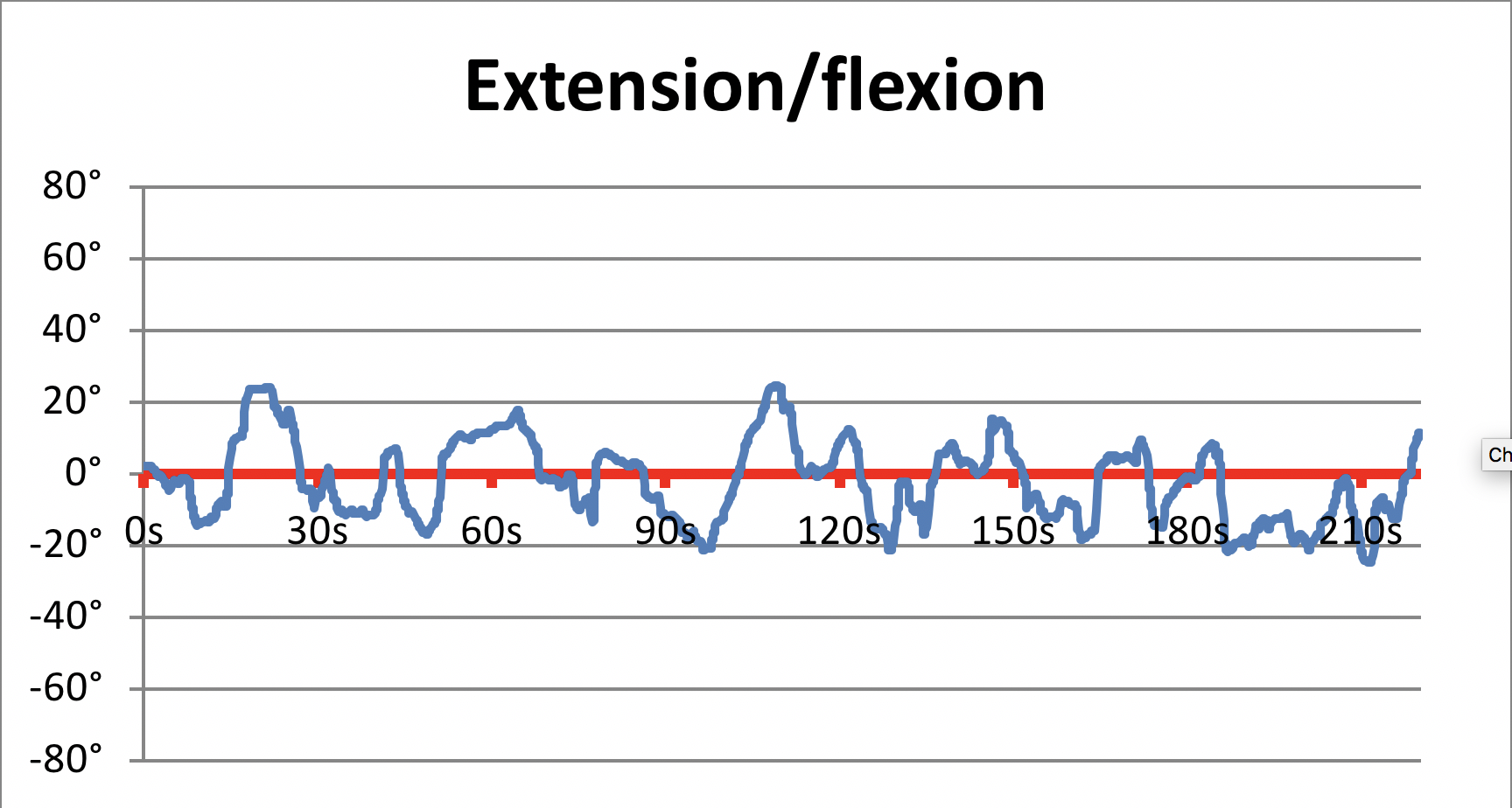}

	\centerline{(b)}
		
	\caption{S3 playing the plane simulator (a) without the bolster and (b) with the bolster.}
	\label{fig:plane}
\end{figure}

We performed four sessions to tune and evaluate our framework. 
%
We asked the therapists to create a level for the plane simulator so that she could simulate an exercise performed in a typical training session. For the flappy bird style game and the skiing game we generated random levels within the therapist requirements. 
Initially, 
patient played without any support but later the bolster was introduced since the therapist noted patients became easily tired. The bolster also improved the tracking of the hand position. Figure~\ref{fig:plane} show the plots of wrist extension/flexion (a) without bolster and (b) with the bolster. As can be note the trends are similar but apart from reducing the fatigue the bolster also helped the patients performing more stable and cleaner movements.

\begin{figure}
	\includegraphics[width=\columnwidth]{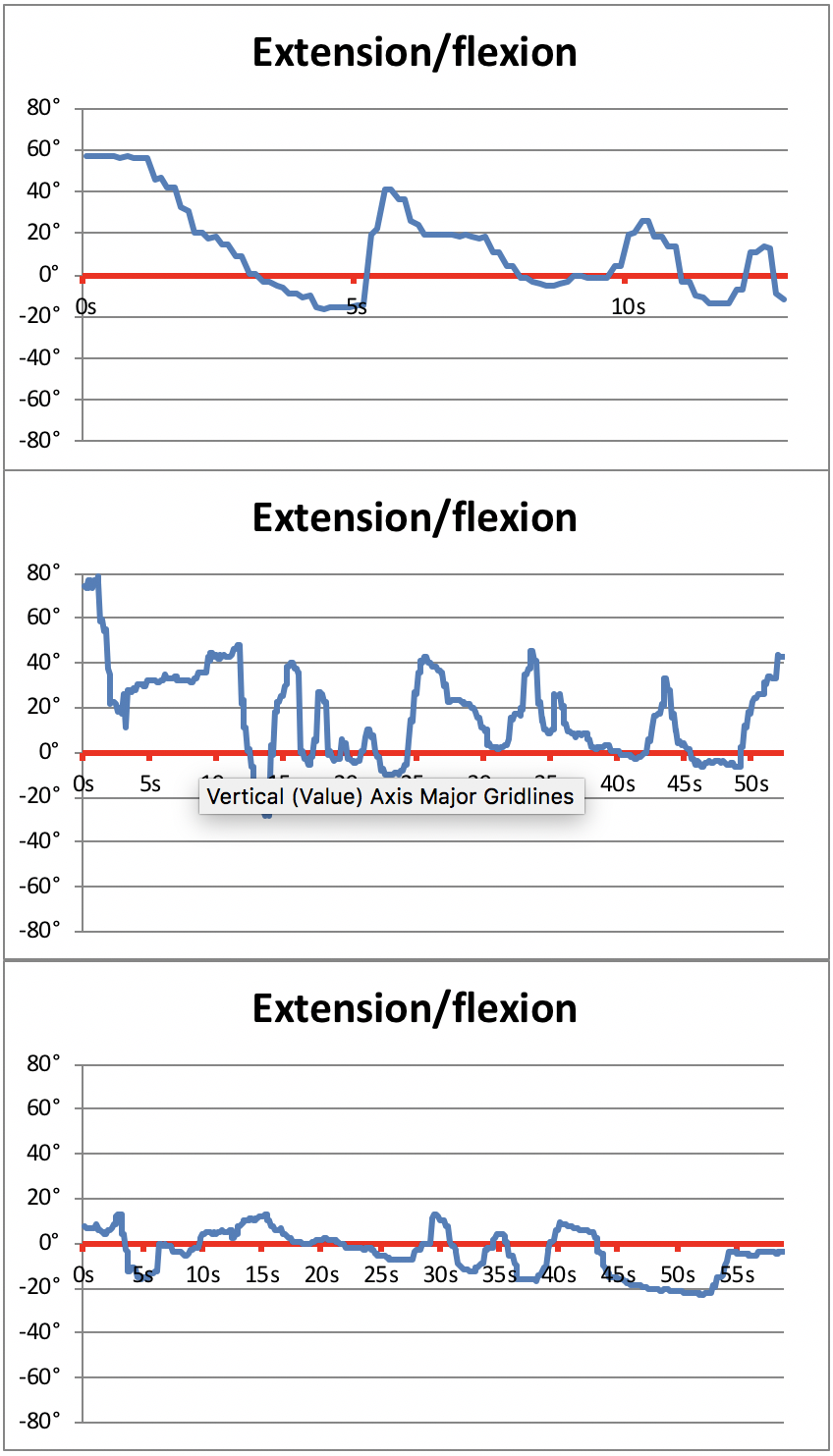}
	\caption{S3 playing the flappy-bird-like game in continuous mode.}
	\label{fig:s2progress}
\end{figure}

The games are quite intuitive and in fact the patients just needed a few games to master the interaction and perform adequate movements 
	coherent to the rehabilitation protocol.
Figure \ref{fig:s2progress} show the plots of wrist extension/flexion for subject S3 over three different games and as can be noted there is a significant improvement over the three runs and the movement become smoother in every run and the games last longer.

\begin{figure}

	\includegraphics[width=\columnwidth]{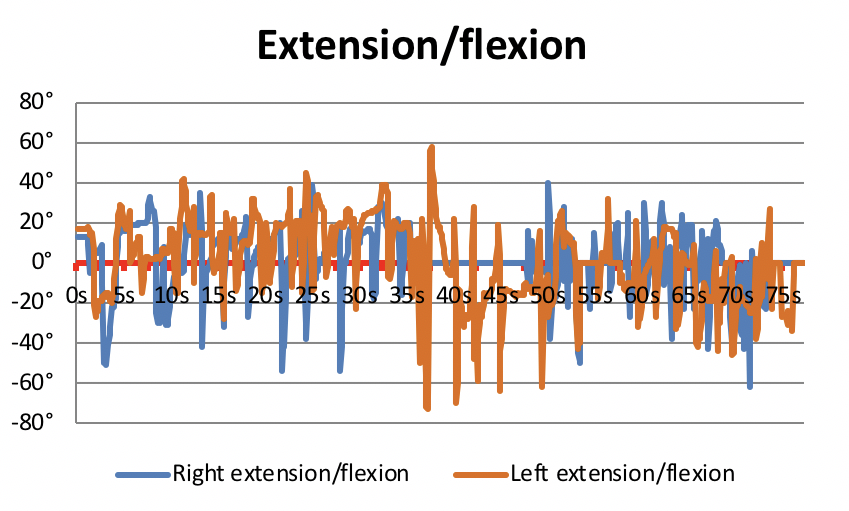}
	
	\centerline{(a)}

	\includegraphics[width=\columnwidth]{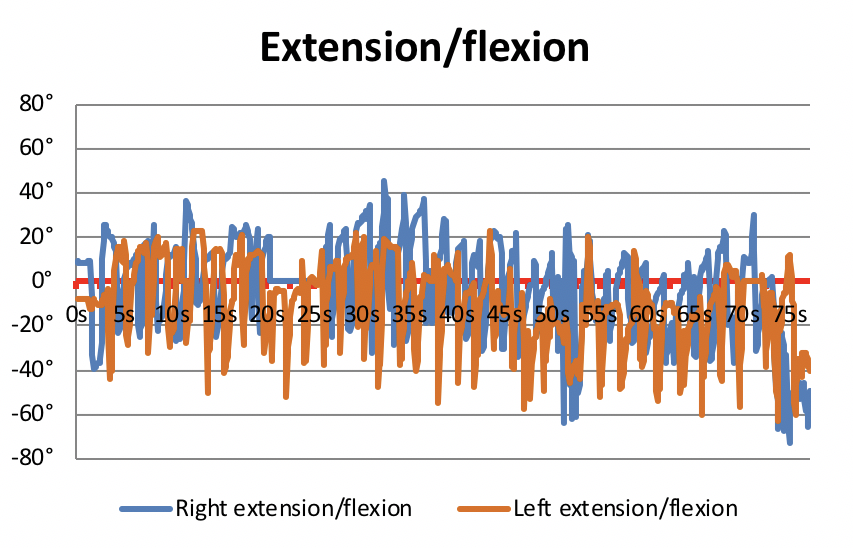}
	
	\centerline{(b)}

\caption{S4 playing the rhythm game in its early version (a) and after it was tuned and the patient had played more games  (b).}
	\label{fig:s4rhythm}
\end{figure}

The rhythm game is more difficult and required some tuning during the sessions. The subjects told us that the game was too complex and asked for simpler levels to play. Figure~\ref{fig:s4rhythm}a reports the wrist extension/flexion for subject S4 during an initial session. The central spikes and the fewer oscillations recorded 
	 suggest that the player had problems keeping up with the rhythm. For the next sessions, we created simpler levels with
	fewer notes to catch (and thus wrist flexion/extensions) and we also increased the distance between two buttons on the same line.
Figure~\ref{fig:s4rhythm}b reports the wrist extension/flexion for subject S4 during a later session. As can be noted the movements appears less abrupt
	and smooth. The subject appeared more confident in the later games as also suggested by the increased score (680 in the initial session and around 800 in the later ones).

During the four sessions we received much feedback from the patients and the therapists that both helped improving the gameplay and tuning
	the difficulty. In general the feedback was very positive. Both therapists and patients liked the scoring system that added an additional challenge 
		among the patients involved. 
Therapists noted that the subjects enjoyed the exercises much more and it did not seem like they were doing  rehabilitation. 
While an exercise performed in the usual setup would raise complaints by the young patients (because they were bored or tired), 
	the same exercises performed using the video game resulted to be engaging and patients never complained.
%
%
The level of focus required by the game, the victory goal and the level of entertainment created by the game itself and by the presence of the other subjects, exceeded the fatigue induced by performing the exercise. 
The skiing game was considered more difficult than others and less intuitive. Thus, we had to tune it, as we did for the rhythm game, by widening the gap between the flags, so that is more easy to pass through them, and widening the track in order to perform a more fluid exercise.
The therapists were satisfied by the replay mode. They really appreciated the possibility to watch all the performed exercise. 

%

%
\section{Conclusion}
The goal of our work was the design of rehabilitation games that could help the patients affected by Juvenile Idiopathic Arthritis 
	performing their physical therapy. 
We wanted to create games that could be both useful and entertaining, that would encourage patients to perform their exercises
	and to maintain their motivation. 
Since JIA mainly targets children and teens, 
	we developed four games with simple mechanics 
	that could appeal boys and girls of a wide age range. 
The games are part of an integrated framework that also provides
	(i) a tuning module, that let the therapists specify the parameter of the exercises;
	(ii) a tracking module that records all the actions performed by the patients;
	(iii) a replay module that let therapists review a session in its entirety;
	(iv) a creation module to let therapists create new exercises manually or through
		a set of constraints.
We performed a preliminary validation with six subjects, that are following 
	a rehabilitation protocol at the clinic, under the supervision of the therapists.
We received positive feedback both from the patients and the therapists. 
The patients liked the games and suggested us some additional changes to make them more appealing. 
The therapists were satisfied with our design and they liked that the games were intuitive and easy 
	to play so that patients did not need much training. 
They also liked the recording capabilities of the system 
	and the possibility of replaying entire sessions.
During the experimental sessions we received also suggestions for future work like for instance 
	a game mode for the rhythm game focused on extension movements, the possibility to 
	use a webcam to let therapists check live patients exercises remotely, 
	and the extension of the framework to lower limbs (ankles and knees).

%

\ifCLASSOPTIONcaptionsoff
  \newpage
\fi



%


\end{document}